\RequirePackage{ifpdf}
\ifpdf 
\documentclass[pdftex]{sigma}
\else
\documentclass{sigma}
\fi

\begin{document}

\renewcommand{\PaperNumber}{112}

\FirstPageHeading

\renewcommand{\thefootnote}{$\star$}

\ShortArticleName{Deformed Oscillators with Two Double Degeneracies}

\ArticleName{Deformed Oscillators with Two Double (Pairwise) \\ Degeneracies of Energy Levels\footnote{This paper is a contribution to the Proceedings
of the Seventh International Conference ``Symmetry in Nonlinear
Mathematical Physics'' (June 24--30, 2007, Kyiv, Ukraine). The
full collection is available at
\href{http://www.emis.de/journals/SIGMA/symmetry2007.html}{http://www.emis.de/journals/SIGMA/symmetry2007.html}}}

\Author{Alexandre M. GAVRILIK  and Anastasiya P. REBESH}

\AuthorNameForHeading{A.M. Gavrilik and A.P. Rebesh}

\Address{Bogolyubov Institute for Theoretical Physics, 14-b Metrologichna Str., 03680 Kyiv, Ukraine}

\Email{\href{mailto:omgavr@mail.bitp.kiev.ua}{omgavr@mail.bitp.kiev.ua}}

\ArticleDates{Received October 09, 2007, in f\/inal form November
13, 2007; Published online November 22, 2007}

\Abstract{A scheme is proposed which allows to obtain special
$q$-oscillator models whose characteristic feature consists in
possessing two dif\/fering pairs of degenerate energy levels. The
method uses the model of two-parameter deformed $q,\!p$-oscillators
and proceeds via appropriately chosen particular relation between
$p$ and $q$. Dif\/ferent versions of quadratic relations $p=f(q)$ are
utilized for the case which implies two degenerate pairs $E_1=E_2$
and $E_3=E_4$. On the other hand, using one f\/ixed quadratic
relation, we obtain $p$-oscillators with other cases of two pairs of
(pairwise) degenerate energy levels. }


\Classification{81S05; 81R50}

\section{Introduction}

The celebrated model of harmonic oscillator in quantum mechanics is
distinguished as fundamental model applicable to a plenty of real
physical systems. But, when applying the harmonic oscillator for
description of real systems, there is often a substantial
discrepancy between theoretical results and experimental data.
Therefore, to make description of a physical system more successful,
it is natural to explore modif\/ications of the model. For this goal,
several deformed versions of the commutation relations of
conventional harmonic oscillator have been proposed. Probably, the
simplest modif\/ied version of the oscillator is the model introduced
in \cite{Arik-Coon}, namely the Arik--Coon (AK) $q$-deformed
oscillator. Subsequently, other types of deformed oscillators have
been also written out: $q$-oscillator of Biedenharn and Macfarlane
(BM) \cite{Biedenharn,Macfarlane}, the so-called Tamm--Dancof\/f (TD)
deformed oscillator \cite{OKK, CSJ}, the $q,\!p$-oscillators
\cite{Chakrabarti-Jagannathan}, and more general models~\cite{Mizr,Man'ko,Melj}.

No doubt, various deformed oscillators, both known and new ones,
deserve detailed study. In our paper we analyze unusual degeneracy
properties of energy levels of $q,\!p$-oscillators (for real~$q$,~$p$) and obtain, on their base, certain $q$-deformed or $p$-deformed
oscillators possessing the property of simultaneous two degeneracies
within two certain pairs of energy levels.

A ``no-go'' theorem is known in quantum mechanics stating that, in one
dimension, the discrete spectrum of any conventional quantum
mechanical system with quadratic kinetic term and nonsingular
potential $V(x)$ does not admit degeneracy of energy levels
\cite{LL}, see also \cite{Parw} for some examples going beyond the
statement. However, there certainly exist more general, than
conventional, systems of quantum mechanics which possess such a
property as degeneracy of some energy levels, even in the absence of
any obvious symmetry (we mean so-called ``accidental'' degeneracy).
Deformed oscillators ($q$-oscillators, $q,\!p$-oscillators) can
supply large class of such systems. As recently demonstrated in
\cite{GR1}, the TD deformed oscillator is just such a special model
which possesses ``accidental'' double degeneracy within a pair of
energy levels, at appropriately f\/ixed value of $q$. Note that the TD
oscillator is contained as a reduced $q=p$ one-parameter case in the
two-parameter generalized family of $q,\!p$-oscillators. What
concerns the $q,\!p$-oscillator, it can be proven \cite{GR2} that,
for appropriate $q$ and $p$, double degeneracy within the
corresponding pair of energy levels takes place. Moreover, starting
from the two-parameter deformed oscillators, a plenty of
$q$-deformed or $p$-deformed oscillators can be derived which
manifest the property, completely analogous to the above mentioned
accidental double degeneracy of the TD oscillator, occurring at
def\/inite values of $q$ or $p$.

The aim of our paper is to demonstrate the existence of special
$q$-oscillator models possessing two double degeneracies (occurring
within two dif\/ferent pairs) of energy levels, namely to describe a
procedure that allows to obtain such $q$-oscillator models from the
two-parameter family of $q,\!p$-oscillators. In most of the
considered particular cases we also show graphically the shape of
the energy as  function of the quantum number $n$ at f\/ixed
deformation parameter(s).

The plan of the paper is the following. In Section~\ref{section2} we recall
necessary facts on the $q,\!p$-oscillator and mention the property of TD oscillators to have double degeneracy. Other
particular examples of $q$-oscillators inferred using various
functional dependencies $p=f(q)$, are indicated in Section \ref{section3} were
dif\/ferent types (pairs) of degenerate energy levels are considered.
In Section~\ref{section4} we examine an interesting case of occurrence of two
double degeneracies of particular pairs of energy levels of
$q,\!p$-oscillators for specially f\/ixed values of~$q$ and~$p$.

For each case of two double degeneracies, a graphical picture
illustrating explicit dependence of energy on the quantum number $n$
is given. Last section is devoted to conclusions.

\section[Necessary facts concerning $q,\!p$-oscillator]{Necessary facts concerning $\boldsymbol{q,\!p}$-oscillator}
\label{section2}

General two-parameter family of $q,\!p$-deformed oscillators is
def\/ined by the relations            \cite{Chakrabarti-Jagannathan}
\begin{gather}
aa^+-qa^+a=p^N, \qquad  aa^+-pa^+a=q^N ,\label{eq1}
\\
 [N,a]=-a ,  \qquad  [N,a^+]=a^+.\label{eq2}
\end{gather}
From the two relations in \eqref{eq1} the main formulas (invariant under
$p\leftrightarrow q$) do follow:
\begin{gather}
a^+a=[\![N]\!]_{q,p} , \qquad aa^+=[\![N+1]\!]_{q,p} , \qquad
[\![X]\!]_{q,p}\equiv \frac{q^X-p^X}{q-p} .\label{eq3}
\end{gather}
We take the Hamiltonian in the form analogous to usual quantum
harmonic oscillator, i.e.,
\begin{gather}
H=\frac{\hbar\omega}{2}(aa^+ + a^+a)  .\label{eq4}
\end{gather}
From now on, $\hbar\omega=1$ is put for simplicity. The
$q,\!p$-analogue of the Fock space, with the vacuum state
$|0\rangle$ obeying $a|0\rangle=0$, is adopted. Then,
\begin{gather}\label{eq5}
\quad |n\rangle=\frac{(a^+)^n}{\sqrt{[\![n]\!]_{q,p}\,!}}|0\rangle,
\qquad N|n\rangle=n|n\rangle
\end{gather}
where $[\![n]\!]_{q,p}!=[\![n]\!]_{q,p}[\![n-1]\!]_{q,p}\cdots [\![2]\!]_{q,p}[\![1]\!]_{q,p}$,
and the $q,\!p$-brackets are def\/ined in \eqref{eq3}. The creation and annihilation
operators act by the formulas
\begin{gather}\label{eq6}
a|n\rangle=\sqrt{[\![n]\!]_{q,p}}|n-1\rangle, \qquad
a^+|n\rangle=\sqrt{[\![n+1]\!]_{q,p}}|n+1\rangle.
\end{gather}
From \eqref{eq4}--\eqref{eq6}, the spectrum $H|n\rangle=E_n|n\rangle$ of the Hamiltonian reads:
\begin{gather}\label{eq7}
E_n\equiv
E_{q,p}(n)=\frac{1}{2}\left(\frac{q^{n+1}-p^{n+1}}{q-p}+\frac{q^n-p^n}{q-p}\right)
=\frac{1}{2}\left(\sum^n_{r=0}q^r p^{n-r}+\sum^{n-1}_{s=0}q^s
p^{n-1-s} \right) .
\end{gather}

At $q\rightarrow 1$ and $p\rightarrow 1$ one recovers the familiar
result $E_n=n+\frac{1}{2}$ as it should. Moreover, at $n=0$ we have
$E_n=\frac{1}{2}$ regardless of the values of $q$ and $p$.

For real values of $q$ and $p$ belonging to the interval $(0,1]$, we
study the issue of ``accidental'' degeneracy of energy levels. The
fact of double degeneracies of certain energy levels was f\/irst
revealed in the paper~\cite{GR1}    where we
examined the degeneracies of various pairs of energy levels of TD
type $q$-oscillator at the corresponding values of the parameter~$q$. In~\cite{GR2},
it is proved that $q,\!p$-oscillators as well exhibit double
degeneracy of energy levels. For instance, it is possible to realize
such degeneracy of energy levels as $E_2=E_3$, or $E_0=E_4$, or
others. In that paper, we dealt with real values of $q$ and $p$
belonging to the interval $(0,1]$.

Since our present paper is focused on the study of {\it two}
double degeneracies of energy levels, the easiest way to achieve
this goal is to start with two-parameter $q,\!p$-oscillators and
then reduce them in a def\/inite way to appropriate one-parameter
$q$-oscillator or $p$-oscillator.

\section{Two double (pairwise) degeneracies of energy levels}\label{section3}

By exploiting the $q$-dependence of energy spectrum of various
$q$-deformed oscillators we can f\/ind that simultaneous degeneracy of
more then one pair of energy levels is possible. In this section, we
develop a procedure for obtaining those special $q$- (or $p$-)
oscillators which possess the unusual property of {\it two double
degeneracies} (within each of two pairs) of energy levels.

So, let us take the $q,\!p$-oscillators def\/ined in \eqref{eq1}--\eqref{eq3} as our
playground. In our whole treatment, $q$ and $p$ attain real values
from the interval $(0,1]$. Obviously, the limit $q\to 1$, $p\to 1$
recovers usual quantum oscillator.

It can be proved that the relation $E_n-E_{n+k}=0$, with $n$ and $k$
being some f\/ixed non-negative integers, determines a continuous
curve in the $(q,p)$-plane. The set of all such curves is naturally
divided into two sets (families):
\begin{gather}\label{eq8}
1) \ \ n\neq 0 , \ \ k\geq 1 \ , \qquad 2) \ \ n=0  , \
\ k\geq 2  .
\end{gather}
The curves from the {\it first family} do not intersect with each
other except for the two points $(1,0)$ and $(0,1)$ which are
nothing but the endpoints for each curve. The curves from the {\it
second family} have no common points; as $k$ grows from~2 to
inf\/inity, their endpoints move along the both axes, starting from
the value $(\sqrt{5}-1)/2$ at $k=2$ and approaching at
asymptotically large~$k$ to~1, on each axis. Both the f\/irst and the
second families in~\eqref{eq8} share the property that the sequence of
curves which correspond to growing integer indices $n+k$ and $k$
respectively, have the tendency of squeezing into the upper right
corner given by the vertex $(1,1)$ and formed by the straight lines
$q=1$, $p=1$.

Now we go over to main point of our scheme: we shall apply dif\/ferent
appropriate choices of functional dependence $p=f(q)$. This way,
from general two-parameter $q,\!p$-oscillator we can obtain
various versions of one-parameter oscillator. The particular case
studied in \cite{GR1} is nothing but the $p=q$ reduction (TD
oscillator), and in this paper we exploit other special dependencies
$p=f(q)$ which can lead to the desired degeneracy properties.
Besides, we often visualize the explicit energy spectrum, with the
occurring degeneracies, of those one-parameter oscillators that are
obtained by exploring a particular choice of $p=f(q)$.

\subsection[$q$-oscillator resulting from quadratic (parabolic) relation $p=f(q)$]{$\boldsymbol{q}$-oscillator resulting from quadratic (parabolic) relation $\boldsymbol{p=f(q)}$}\label{section3.1}

(i) Let us take
\begin{gather}\label{eq9}
p=\alpha (q-\beta)^2 + \gamma ,
\end{gather}
where $\alpha$, $\beta$, $\gamma$ are the parameters of the curve.
By inserting this in basic formulas for $q,\!p$-oscillators, a
particular one-parameter set of new deformed oscillators is
obtained. The resulting $q$- (or $p$-) oscillators will {\it possess
the property of two double degeneracies of energy levels}, if the
simultaneous degeneracy within each of, e.g., the following two
pairs does occur:
\begin{gather}\label{eq10}
E_1=E_2  , \qquad E_3=E_4
\end{gather}
(note that any other concrete two pairs of energy levels, except for
$E_0=E_1$, can be chosen).

In order that such two-fold double degeneracy \eqref{eq10} of energies would
emerge, the curve corresponding to relation \eqref{eq9} whose ef\/fect
consists in reducing the $q,\!p$-oscillator to some one-parameter
$q$- or $p$-oscillator, should cross the curves $E_2-E_1=0$ and
$E_4-E_3=0$ at the points A$(q_1,p_1)$ and B$(q_2,p_2)$
respectively, where $p_1=p_2=p_0$. Here, the coordinates $q_1$ and
$q_2$ are evaluated, at f\/ixed $p_0$, from solving the equations
\eqref{eq10}. Moreover, the curve \eqref{eq9} should contain the point $(1,1)$ in
order that $p(1)=1$ since, obviously, each $q$- or $p$- oscillator
at $q\rightarrow 1$ and $p\rightarrow 1$ must turn into the usual
harmonic oscillator.

Then, the following system of equations  (recall, $p_1=p_2=p_0$)
\begin{gather}
 \alpha(q_1 - \beta)^2 + \gamma =p_1  , \qquad
 \alpha(q_2 - \beta)^2 + \gamma =p_2 ,  \qquad
 \alpha(1 - \beta)^2 + \gamma =1  ,\label{eq11}
\end{gather}
is used for f\/inding  $\alpha$, $\beta$, $\gamma$ that specify the curve in \eqref{eq9}.
The solution of the system reads:
\begin{gather}\label{eq12}
\alpha = \frac{1-p_0}{(1-q_1)(1-q_2)} ,\qquad \beta = \frac{q_1+q_2}{2} , \qquad
\gamma = 1+ \frac{(p_0-1)}{(1-q_1)(1-q_2)}\Bigl(1-\frac{q_1-q_2}{2}\Bigr)^2  .
\end{gather}
We put these $\alpha$, $\beta$ and $\gamma$ in the parabolic
relation \eqref{eq9} and get it completely specif\/ied.
   Then, by inserting the inverted relation $q=f^{-1}(p)$ into the
def\/ining formulas \eqref{eq1}--\eqref{eq3} of $q,\!p$-oscillator we derive a
special $p$-oscillator such that at $p=p_0$ the desired two-fold
double degeneracy (i.e., the simultaneous degeneracies \eqref{eq10} of two
dif\/ferent pairs of energy levels) is guaranteed.

To illustrate the obtained pattern of degeneracies let us f\/ix for
example $p=p_0=0.6$. Put this value of $p_0$ into equations in \eqref{eq10}
and solve them for $q_1$ and $q_2$. As result, the numerical values
of $q_1$ and $q_2$ are at our disposal (we give them up to~6
digits). With these $p_0$, $q_1$ and $q_2$ inserted in \eqref{eq12} we f\/ind
$\alpha$, $\beta$ and $\gamma$; the entire set of values is placed
in Table~\ref{table1}.

\begin{table}[t] \centering
\caption{Values of the parameters in equation~\eqref{eq12} corresponding to $p=p_0$.\label{table1}}

\vspace{1mm}

\begin{tabular} {|c|c|c|c|c|c|} 
\hline
 {} & {$q_1$} & {$q_2$} & {$\alpha$} & {$\beta$} & {$\gamma$}  \\
\hline
$p_0=0.6$  & {0.554400} & {0.900317} & {9.005207} & {0.727359} & {0.330613}  \\
\hline
\end{tabular}
\end{table}

\begin{figure} [t]   
\centering
\begin{minipage}{0.45\textwidth}
 \hspace{2mm}
\includegraphics[angle=-90, width=0.95\textwidth]{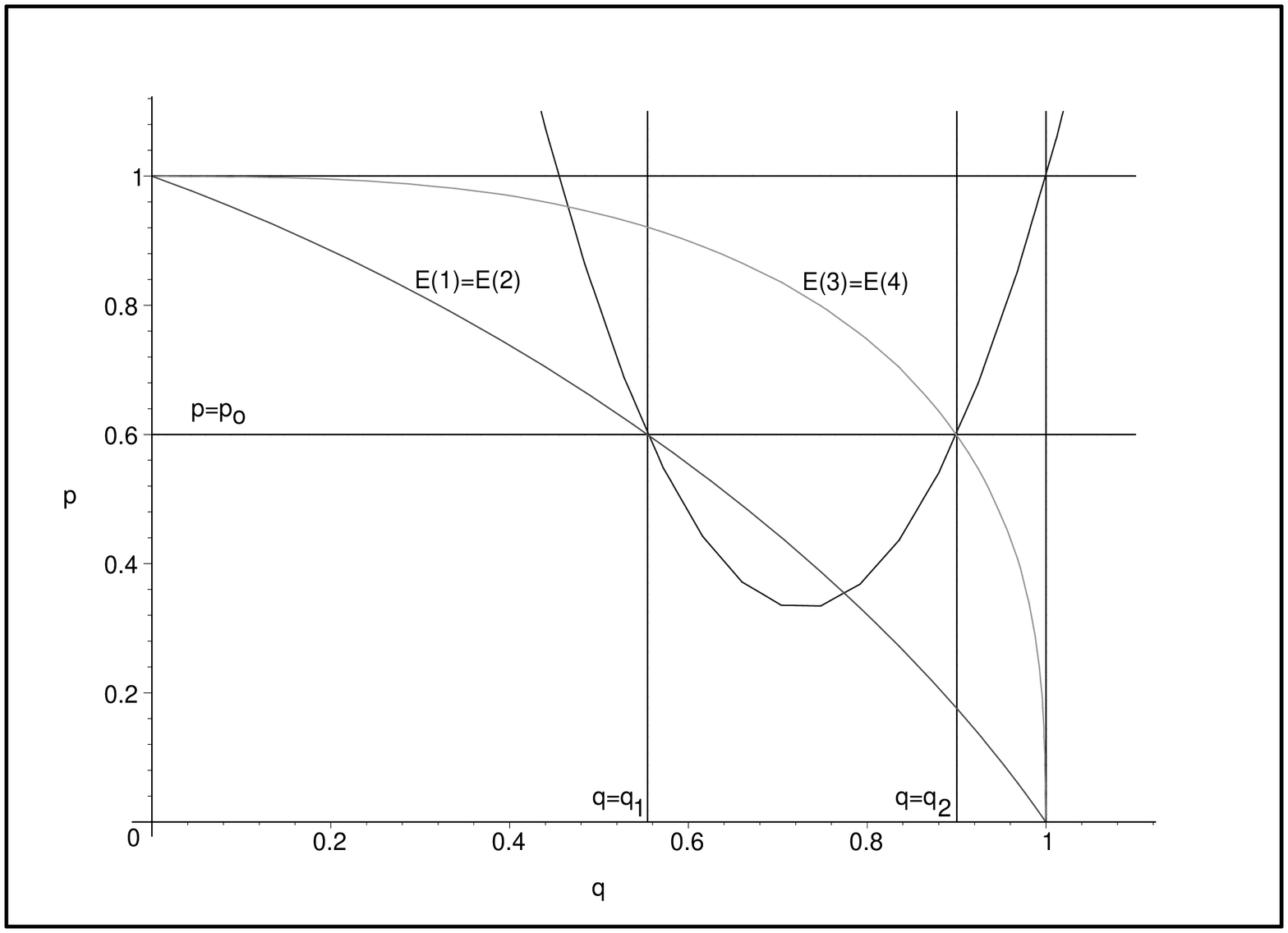}
 \end{minipage}
 \hfil
 \begin{minipage}{0.45\textwidth}
 \includegraphics[angle=-90, width=0.95\textwidth]{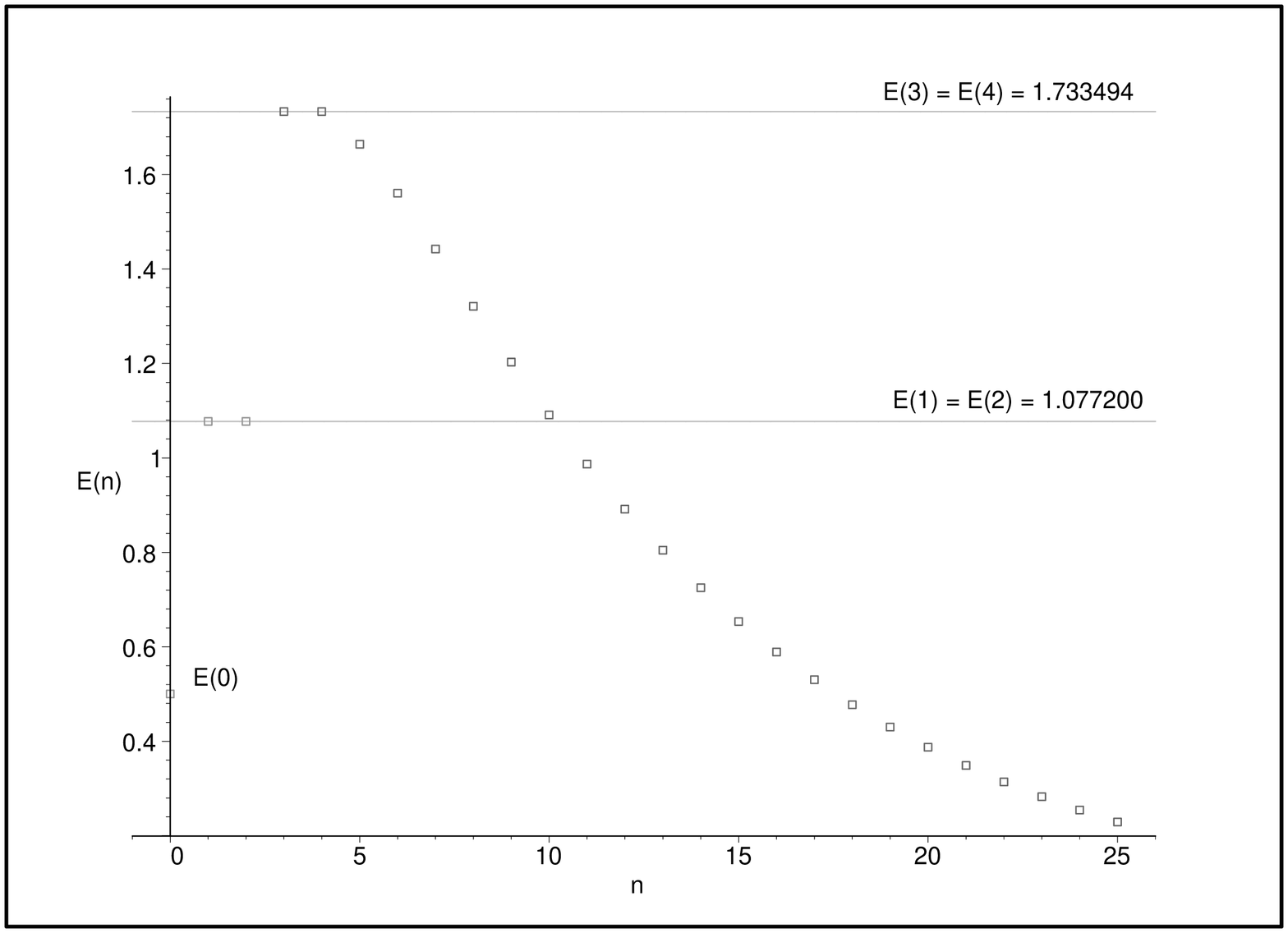}
\end{minipage}
\hspace{-6mm} \caption{{\em Left}: Two double degeneracies of energy
levels $E_1=E_2$ and $E_3=E_4$ occurring at the f\/ixed $p\equiv
p_0=0.6$. The parabolic curve $p=f(q)$ is given by \eqref{eq9} with
$\alpha$, 
$\beta $ 
and $\gamma $ taken from Table~\ref{table1} ($q_1$, $q_2$ are there, too). The
parabola goes through the points $(q_1,0.6)$,
                                     $(q_2,0.6)$
and $(1,1)$ as it should.
{\em Right}: Energy spectrum $E_{p}(n)$, at $p=0.6$, given by the
expression with ``$+$'' or ``$-$''  in \eqref{eq14} respectively for $n=0,1,2$
or $n=3,4,5,\ldots$ (see main text). The two degeneracies $E_1=E_2$
and $E_3=E_4$ are evident.}\label{fig1}
\end{figure}

In  Fig.~\ref{fig1} (left) the resulting picture of two-fold double
degeneracy, see \eqref{eq10}, is shown.

Note that the admissible values of $p$ for the $p$-oscillator  
are such that $p \in [p_{\min},1]$ where $p_{\min}$ is the minimal
value of the function \eqref{eq9} whose image is the parabola. In case at
hands,
\begin{gather*}
p_{\min}=\gamma = 0.330613  .
\end{gather*}

It is of interest to visualize the behavior of energy $E_n=E(n)$ as
function of $n$ for the so obtained $p$-oscillator. Recall that the
specif\/ied values $q_1$, $q_2$, $\alpha$, $\beta$, $\gamma$ are
determined at the f\/ixed $p_0$. Since a one-parameter deformed
oscillator possesses, by construction, the desired property of {\it
two double degeneracies occurring at the same $p=p_0$}, we should
end up (not with the $q$-, but) with the $p$-oscillator. Therefore,
we invert the considered parabolic relation: taking $q=f^{-1}(p)$,
we substitute this in equation~\eqref{eq7}.
 Since the inverse $f^{-1}(p)$, given as
\begin{gather}\label{eq14}
q=\pm \sqrt{\frac{p-\gamma}{\alpha}}+\beta ,
\end{gather}
contains the ``$\pm$'' type ambiguity through pre-factor of square
root (accordingly, we have right or left branch of the parabola in
Fig.~\ref{fig1} (left)), one should be careful when inserting this~$f^{-1}(p)$ into~$E_{q,p}(n)$ in \eqref{eq7}: the two signs result in two
expressions for the one-parameter $E_{p}(n)$. Namely, the ``minus''
square root enters the $E_{p}(n)$ for $n=0,1,2$, whereas the ``plus''
sign determines~$E_{p}(n)$ for the rest of levels $n=3,4,5,\ldots$.
The resulting energy spectrum $E_{p}(n)$, at $p=0.6$, with the
occurring two degeneracies $E_1=E_2$ {\it and} $E_3=E_4$ is shown in
Fig.~\ref{fig1} (right).

\subsection[Hyperbolic/elliptic relations $p=f(q)$ and the $p$-oscillators possessing
            two double degeneracies]{Hyperbolic/elliptic relations $\boldsymbol{p=f(q)}$ and the $\boldsymbol{p}$-oscillators\\ possessing
            two double degeneracies}\label{section3.2}

For completeness and also for comparison it is worth to consider,
besides parabolic, the other two cases of quadratic dependence of
$p$ on $q$: the hyperbolic and the elliptic ones.

(ii) Let us take
\begin{gather}\label{eq15}
(p-{\tilde a})^2-{\tilde b}(q-{\tilde c})^2=R^2 .
\end{gather}
For convenience, let $R=1$ for what follows. Again we are able to
design a $p$-oscillator possessing two simultaneous double
degeneracies. Fixing $p$ in equations \eqref{eq10} as, say, $p_0=0.6$ we
solve them to get $q_1=0.554400$,  $q_2=0.900317$. Then we
compose relevant system of three equations analogous to \eqref{eq11}, but
now encoding the fact that the {\it hyperbolic} curve runs through
the points A$(q_1,p_1)$, B$(q_2,p_2)$, C$(1,1)$, where
$p_1=p_2=p_0$.
  Solving this system, we f\/ind the parameters ${\tilde a}$, ${\tilde b}$,
${\tilde c}$ in \eqref{eq15}. The resulting set of numerical values is
collected in Table~\ref{table2}.

\begin{table}[t] \centering

\caption{Values of the parameters in equation \eqref{eq15} corresponding to $p=p_0$.}\label{table2}

\vspace{1mm}

\begin{tabular} {|c|c|c|c|c|c|} 
\hline
\tsep{0.5ex} {} & {$q_1$} & {$q_2$} & {$\tilde a$} & {$\tilde b$} & {$\tilde c$}  \\
\hline
$p_0=0.6$  & {0.554400} & {0.900317} & $-0.755814$ & {28.020856} & {0.727359}  \\
\hline
\end{tabular}
\end{table}

\begin{figure}[t]   
\centering
\begin{minipage}{0.45\textwidth}
 \hspace{2mm}
\includegraphics[angle=-90, width=0.95\textwidth]{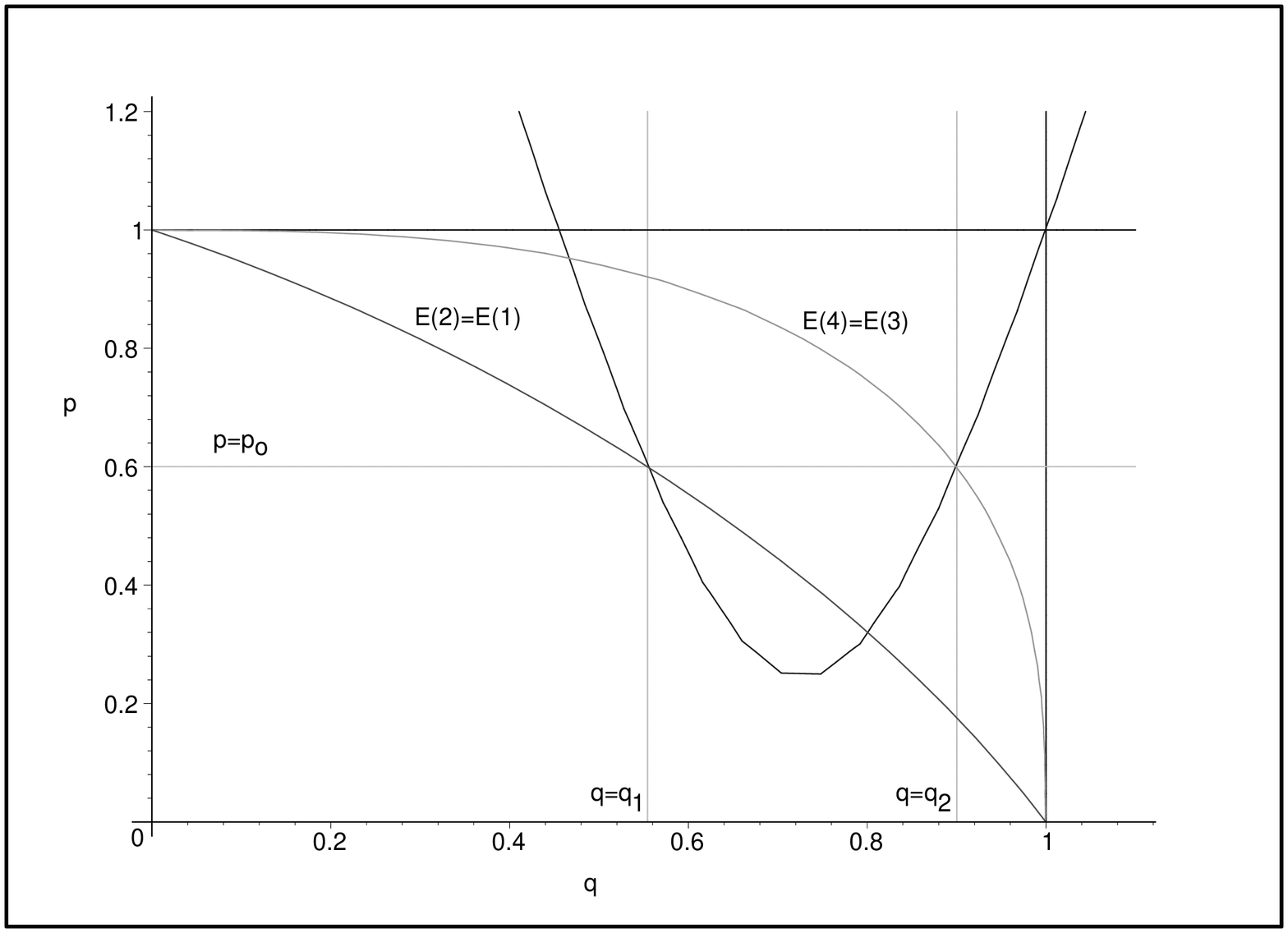}
 \end{minipage}
 \hfil
 \begin{minipage}{0.45\textwidth}
 \includegraphics[angle=-90, width=0.95\textwidth]{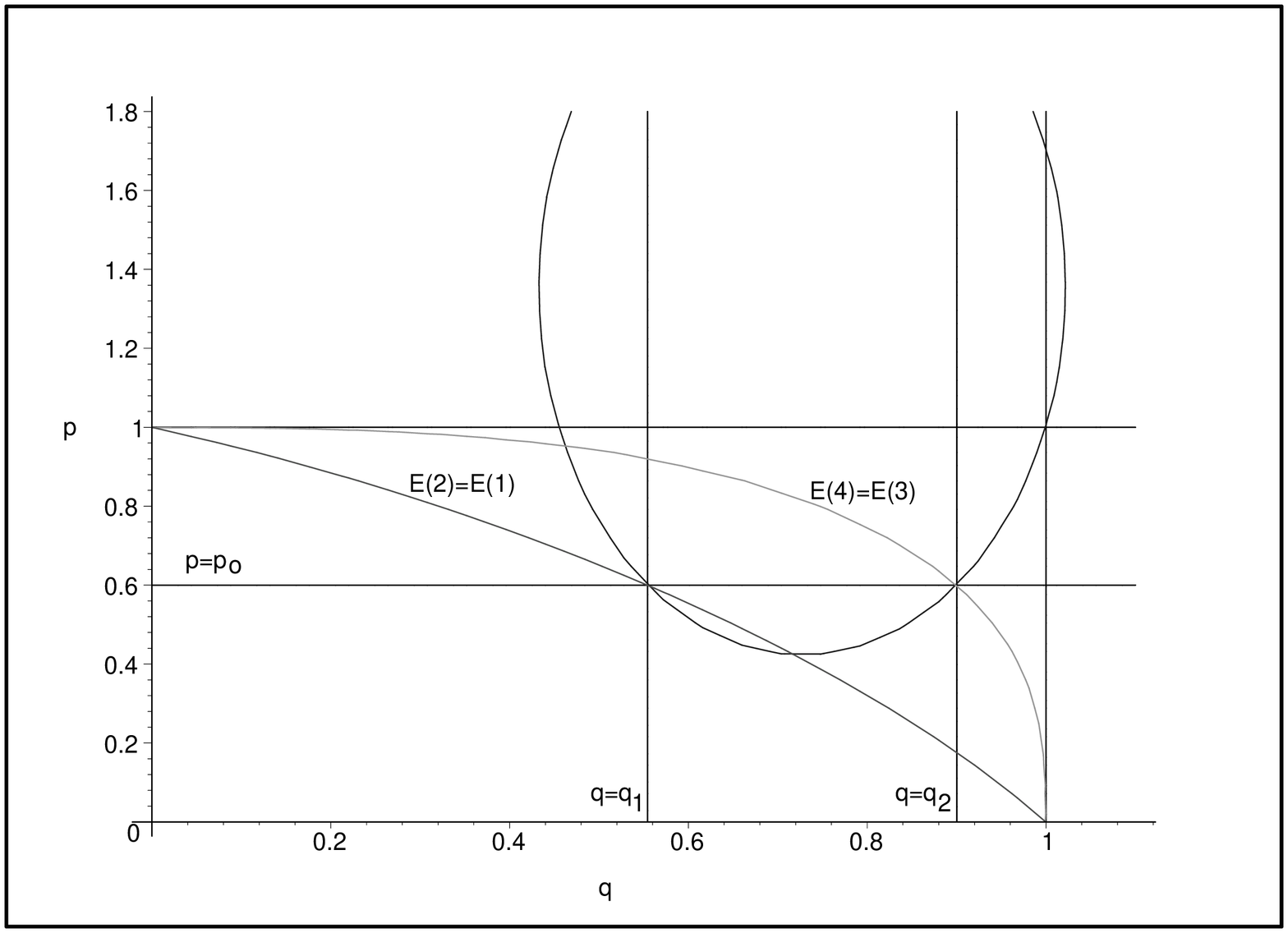}
 \hspace{-10mm}
\end{minipage}
\caption{{\em Left}: Same as on Fig.~\ref{fig1} (left), but now
using hyperbolic relation \eqref{eq15} with $\tilde{a}$, $\tilde{b}$ and
$\tilde{c}$ given in Table~\ref{table2}. As seen, the curve goes through the
points ($q_1$,~0.6), ($q_2$,~0.6) and (1,1).
{\em Right}: Same as on the left panel, but here the elliptic
relation \eqref{eq17} is used, with $\varepsilon =0.1$ and $\mu$, $\nu$,
$\rho$ given in Table \ref{table3}. }\label{fig2}
\end{figure}

With the obtained data we arrive at the picture shown in Fig.~\ref{fig2}
(left). There, the two-fold double degeneracy is manifest (now got
due to {\it hyperbolic} relation $p=f(q)$;  it is the two branches
of hyperbola that again do the job). For clarity, the value
$p_0=0.6$ is also indicated.

The minimal admissible positive value $p_{\min}$ for the
$p$-oscillator obtained by inserting $q=f^{-1}(p)$ of this
hyperbolic relation is found from{\samepage
\begin{gather*}
(p_{\min}-\tilde a)^2-\tilde b(Q-\tilde c)^2=1^2,
\end{gather*}
where $Q=\frac{q_1+q_2}{2}$; for ${q_1,q_2}$ and {$\tilde a$},
{$\tilde b$}, {$\tilde c$} see Table~\ref{table2}. With these values,
$p_{\min}=0.244186$.}

The points located below $p=p_{\min}$ can't belong to the chosen
curve and thus are not admitted for the obtained $p$-oscillator; for
those points $(q,p)$ for which $p<p_{\min}$, we have to construct
another one-parameter oscillator, using the relation $p=f(q)$
compatible with f\/ixed $p_0$ and, then, with the corresponding values
$q_1$, $q_2$, {$\tilde a$}, {$\tilde b$}, {$\tilde c$}.

(iii) Now let us take
\begin{gather}\label{eq17}
(q- \mu )^2+\varepsilon (p- \nu )^2= \rho ^2,
\end{gather}
where $ \mu $, $ \nu $, $ \rho $, $ \varepsilon $ ($\varepsilon >0$
for this elliptic case) are the relevant parameters. By solving the
system of equations composed analogously to \eqref{eq11} these parameters
are easily found and for $p_0=0.6$ placed in Table~\ref{table3}.

\begin{table}[t] \centering

\caption{Values of the parameters in
equation~\eqref{eq17} corresponding to $p=p_0$ and $\varepsilon =0.1$.}\label{table3}
\vspace{1mm}

\begin{tabular} {|c|c|c|c|c|c|} 
\hline
 {} & {$q_1$} & {$q_2$} & {$\mu $}   & {$\nu $}    &  {$\rho $} \\
\hline
$p_0=0.6$  & {0.554400} & {0.900317} & {0.727359} & {1.355234} & {0.294877}  \\
\hline
\end{tabular}

\end{table}

The present case based on the elliptic relation $p=f(q)$ in \eqref{eq17} is
illustrated in Fig.~\ref{eq2} (right).

Our goal in this section was to construct  special one-parameter
deformed oscillators which possess the property of two-fold double
degeneracy of some two pairs of energy levels. Here, the two pairs
$E_1=E_2$ {\it and} $E_3=E_4$ where chosen, and we have succeeded to
realize the goal. Namely, starting from the two-parameter
$q,\!p$-oscillators def\/ined in \eqref{eq1}--\eqref{eq3}, by imposing dif\/ferent
types of the quadratic relations $p=f(q)$: parabolic, hyperbolic and
elliptic, we have inferred the corresponding versions of
one-parameter $p$-deformed oscillators, all of them sharing the
property that simultaneously $E_1=E_2$ and $E_3=E_4$.

\section[$p$-oscillators with other pairs of degenerate energy levels]{$\boldsymbol{p}$-oscillators with other pairs of degenerate energy levels}
\label{section4}

The pairs $E_1=E_2$ and $E_3=E_4$ considered in the previous
section, serve as typical examples belonging to the f\/irst family in
\eqref{eq8}. In this section, we derive and explore one-parameter deformed
oscillators with similar property of two-fold double degeneracy, but
for {\it another two pairs of energy levels} being specif\/ied.
Moreover, we will distinguish the following two cases: (a) the both
pairs of energy levels belong to the second family in \eqref{eq8}; (b) the
two pairs are taken from dif\/ferent families (i.e., mixed case).

\subsection[$p$-oscillators from relation $p=f(q)$: other pairs of energy levels]{$\boldsymbol{p}$-oscillators from relation $\boldsymbol{p=f(q)}$: other pairs of energy levels}\label{section4.1}

Here we construct, using only the parabolic relation \eqref{eq9} of $p$ and
$q$, the $p$-deformed oscillator possessing two double degeneracies
for the cases in which the both pairs of energy levels belong to the
f\/irst family in \eqref{eq8}.

That is, following the same procedure as above, let us f\/ix some two
pairs (each one degenerate, as we require) of {\it energy levels
from the second family} in \eqref{eq8}, say,
\begin{gather}\label{eq18}
E_2-E_0=0 , \qquad
E_5-E_0=0  .
\end{gather}
With some f\/ixed $p=p_0$, let it be $p_0=0.4$, we solve the equations
in \eqref{eq18} for $q_1$, $q_2$ and get: $q_1=0.264365$, $q_2=0.721012$.
Taking these into account, from the formulas \eqref{eq12} we f\/ind  $\alpha$,
$\beta$ and $\gamma$ which specify the parabolic curve \eqref{eq9}, see
Table~\ref{table4} for the relevant set of data.

\begin{table}[t] \centering

\caption{Values of the parameters in
equation~\eqref{eq9} corresponding to $p=p_0$ and equation~\eqref{eq18}.}\label{table4}

\vspace{1mm}

\begin{tabular} {|c|c|c|c|c|c|} 
\hline
  {}     &    {$q_1$}  & {$q_2$}   & ${\alpha}$ & {$\beta$} & {$\gamma$}  \\
\hline
$p_0=0.4$ & {0.264365} & {0.721012} & {2.923499} & {0.492688} & {0.247594}  \\
\hline
\end{tabular}
\end{table}

\begin{figure}[t]   
\centering
\begin{minipage}{0.45\textwidth}
 \hspace{2mm}
\includegraphics[angle=-90, width=0.95\textwidth]{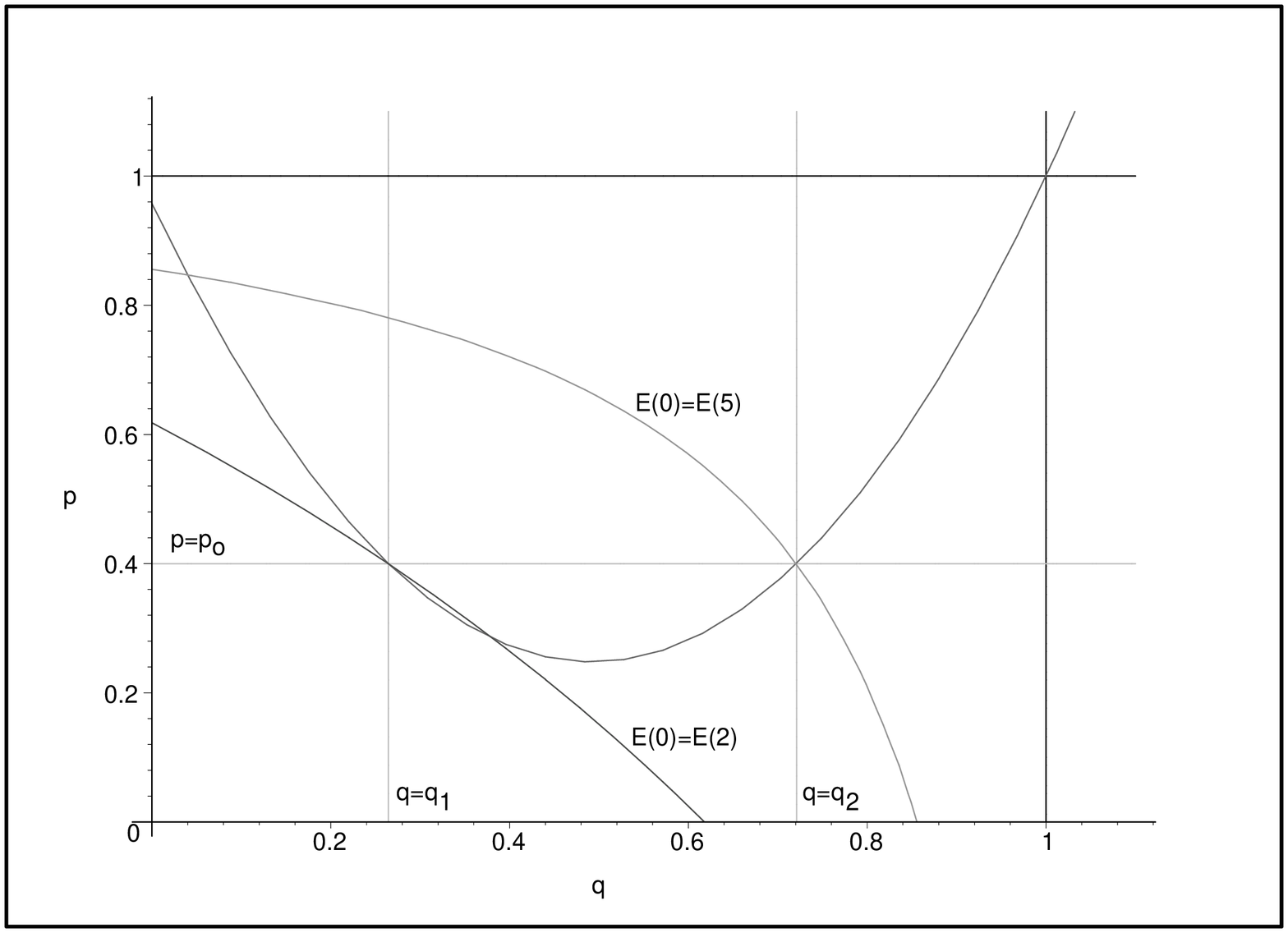}
 \end{minipage}
 \hspace{6mm}
 \begin{minipage}{0.45\textwidth}
 \includegraphics[angle=-90, width=0.95\textwidth]{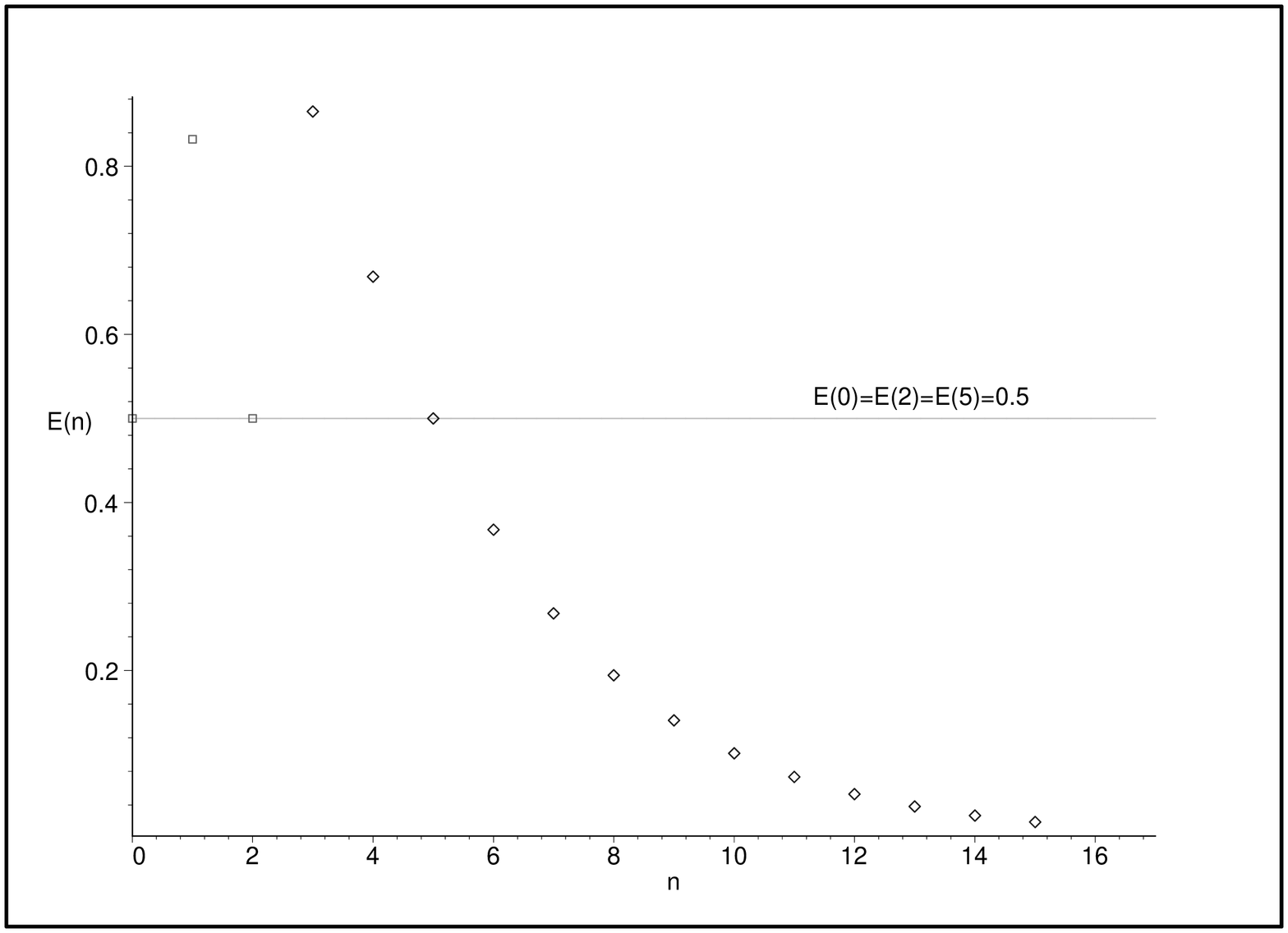}
 \hspace{-10mm}
\end{minipage}

\caption{{\em Left}: Intersection, at $p_0=0.4$, of
the curves $E_2-E_0=0$ and $E_5-E_0=0$ by the parabolic curve \eqref{eq9}
whose parameters $\alpha$, $\beta$ and $\gamma$ are given in Table~\ref{table4}. {\em Right}: Simultaneous degeneracies, at $p=0.4$, of the levels
$E_{2}=E_0$ and $E_5=E_0$ in the energy spectrum of the derived
$p$-oscillator. Two formulas for~$E_p(n)$, one for $n=0,1,2$, the
other for $n=3,4,5,\dots$, realize the ``$\pm$'' dichotomy in \eqref{eq14}.}\label{fig3}
\end{figure}

The resulting $p$-deformed oscillator does possess the prescribed
pattern of degeneracies \eqref{eq18}, and in Fig.~\ref{fig3} (left) this is neatly
seen. The shape of the related energy spectrum $E_p(n)$ is presented
in Fig.~\eqref{eq3} (right). Note the peculiar feature of non-monotonicity of
the energy function which is caused by the fact that, because of the
``$\pm$'' pre-factor of square root in \eqref{eq14}, one should use the two
alternative expressions for $E_p(n)$ when $n=0,1,2$ or when
$n=3,4,5,\dots$.

Let us f\/inally explore the mixed case. We use again the parabolic
relation \eqref{eq9} of $p$ and $q$, and construct the $p$-deformed
oscillator which possesses the two-fold double degeneracy for the
{\it case when the two pairs belong to different families} in \eqref{eq8}.
As a typical example let us take
\begin{gather}\label{eq19}
E_3-E_2=0 , \qquad
E_4-E_0=0  .
\end{gather}
Following the same procedure as above, having f\/ixed $p=p_0$ as,
e.g., $p_0=0.4$, we solve the equations in \eqref{eq19} and obtain:
$q_1=0.640778$, $q_2=0.916515$. Taking these into account, from the
formulas \eqref{eq12} we f\/ind  $\alpha$, $\beta$, $\gamma$ (see Table~\ref{table5}),
which completely specify the parabolic curve~\eqref{eq9} in this case.

\begin{table}[t] \centering

\caption{Values of the parameters in
equation~\eqref{eq9} corresponding to $p=p_0$ and equation~\eqref{eq19}.}\label{table5}

\vspace{1mm}

\begin{tabular} {|c|c|c|c|c|c|} 
\hline
  {}     &    {$q_1$}  & {$q_2$}   & ${\alpha}$ & {$\beta$} & {$\gamma$}  \\
\hline
$p_0=0.4$ & {0.640778} & {0.916515} & {20.006946} & {0.778648} & {0.019714}  \\
\hline
\end{tabular}
\end{table}

\begin{figure}[t]
\centering
\begin{minipage}{0.45\textwidth}
 \hspace{2mm}
\includegraphics[angle=-90, width=0.95\textwidth]{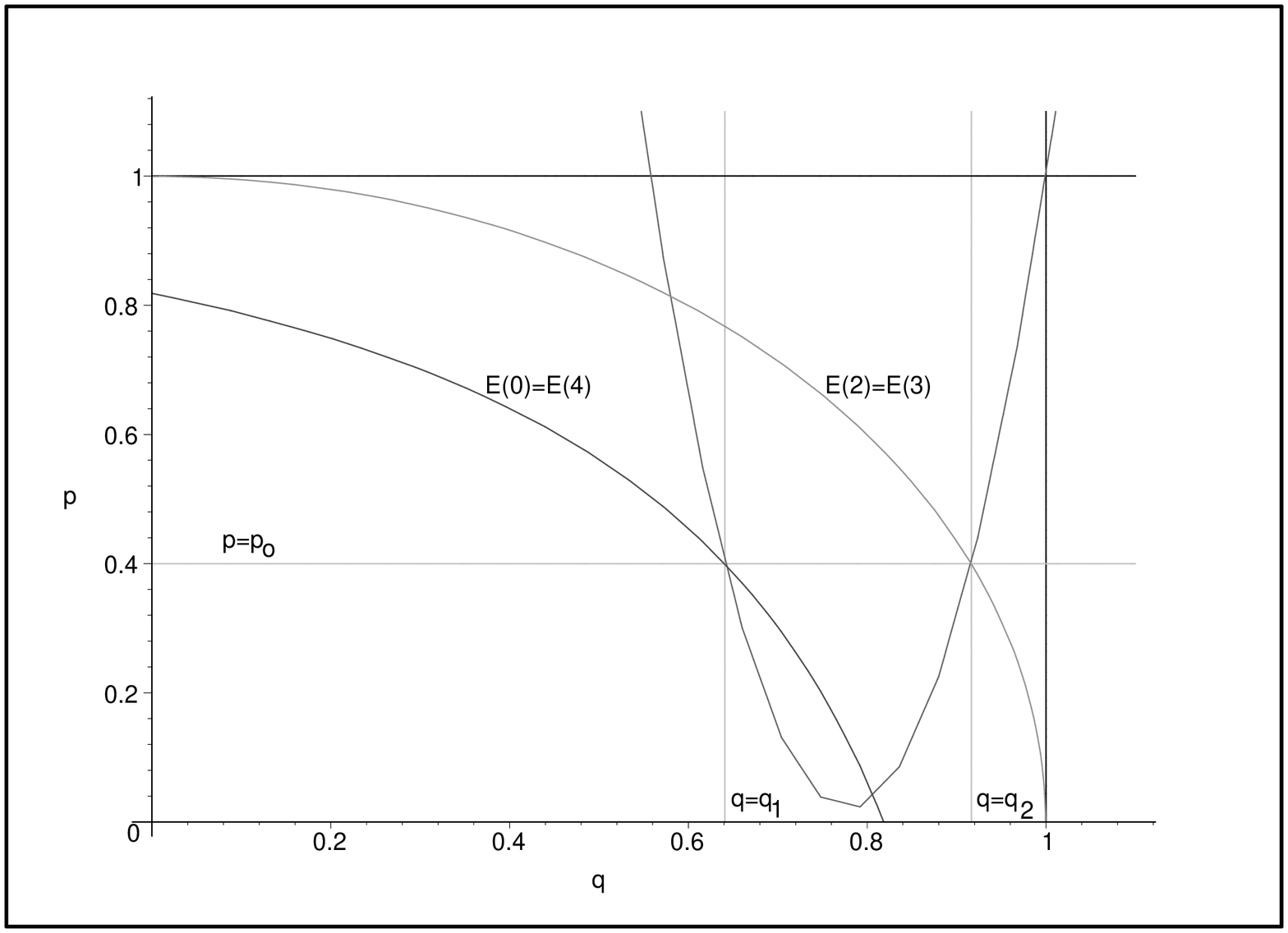}
 \end{minipage}
 \hspace{6mm}
 \begin{minipage}{0.45\textwidth}
 \includegraphics[angle=-90, width=0.95\textwidth]{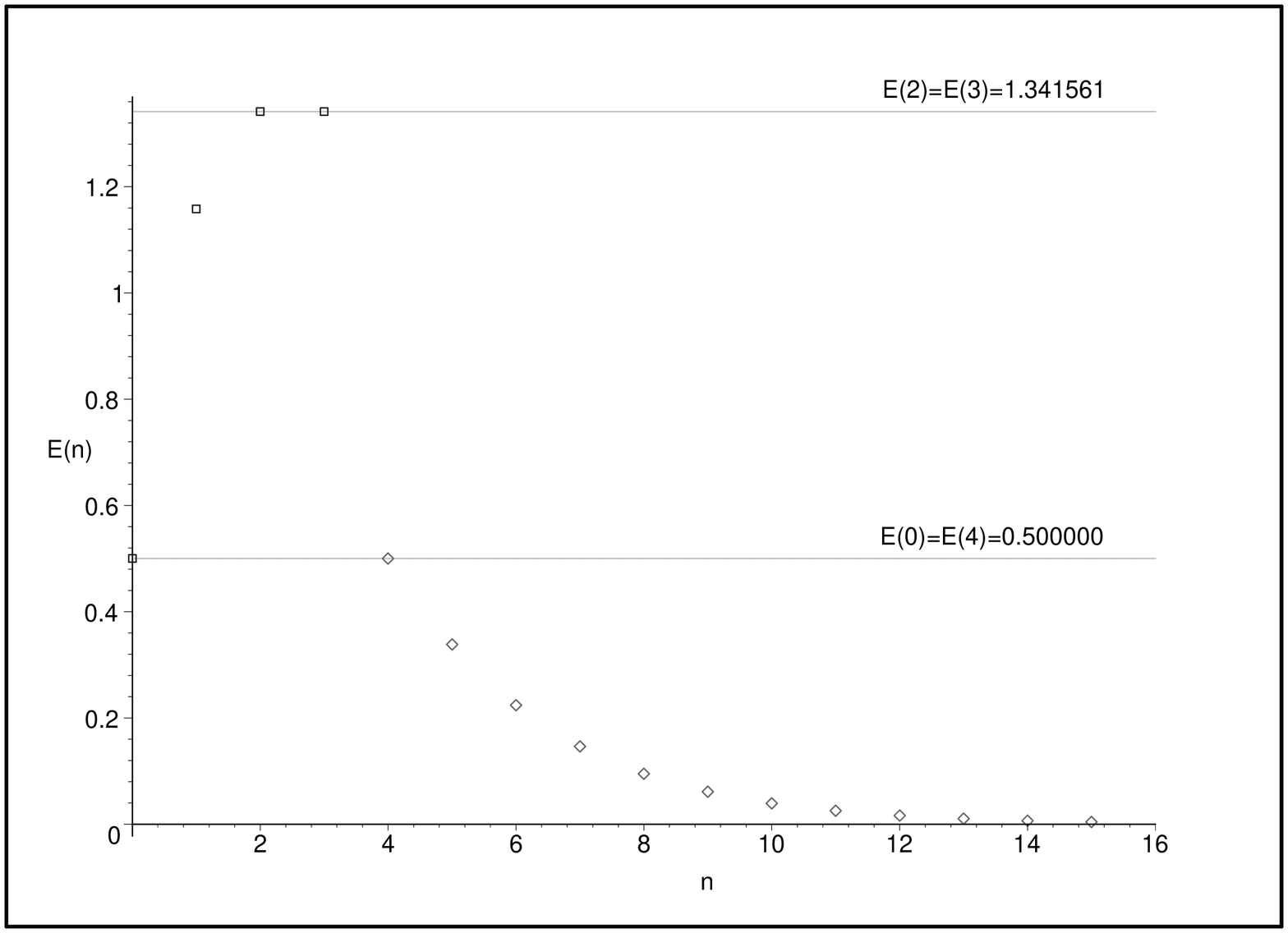}
 \hspace{-10mm}
\end{minipage}

\caption{{\em Left}: Intersection of the curves
$E_4-E_0=0$ and $E_3-E_2=0$ by the parabolic curve \eqref{eq9}, at f\/ixed
$p=p_0=0.4$ and with $q_1$, $q_2$, $\alpha$, $\beta$ and $\gamma$
given in Table~\ref{table5}.
{\em Right}: Energy spectrum at $p=0.4$, manifesting simultaneous
degeneracies of the two pairs of energy levels $E_4=E_0$ and
$E_3=E_2$. }\label{fig4}
\end{figure}

As should, the resulting $p$-deformed oscillator acquires the prescribed pattern \eqref{eq19}
of degeneracies.
   In Fig.~\ref{fig4} (left) this pattern is clearly seen. The shape of the related energy
spectrum~$E_p(n)$ is shown in Fig.~\ref{fig4} (right). Note again the
peculiar feature of non-monotonicity of the energy function caused
by the fact that, because of the ``$\pm$'' pre-factor of square root
in \eqref{eq14}, one should use the proper version of the two alternatives
for $E_p(n)$: one for $n=0,1,2,3$, the other for $n=4,5,\dots$.

\subsection[$q$-oscillator with two double degeneracies from
                linear relation of $p$ and $q$]{$\boldsymbol{q}$-oscillator with two double degeneracies from
                linear relation of $\boldsymbol{p}$ and $\boldsymbol{q}$}\label{section4.2}

Above, we constructed one-parameter oscillators on the base of the
two-parameter $q,\!p$-oscillator by imposing specif\/ied quadratic
relation $p=f(q)$. It turns out however that two-fold double
degeneracy can be gained {\it at the two-parameter level}
($q,\!p$-deformed level) more simply, for two def\/inite (not
arbitrary) pairs of energy levels. This is due to the fact that
certain two curves belonging to dif\/ferent families of
$E_{n+k}-E_n=0$ in \eqref{eq8} may merely intersect each other. Indeed, this
occurs if one curve is given by $E_k-E_0=0$ from the second family
in \eqref{eq8} with $k$ being high enough, and the other curve taken from
the set $E_{n+l}-E_n=0$, $n\neq 0$, with an appropriate rather small
$n$ and $l$.

In such cases it may occur that the specif\/ied curves have
simultaneously two dif\/ferent (mutually $p\leftrightarrow q$
symmetric) points of intersection.
 This implies the following:
simultaneous degeneracies within two certain pairs of energy levels
takes place at two distinct couples of f\/ixed  $q$, $p$ values.
\begin{figure}[t]   
\centering
\begin{minipage}{0.45\textwidth}
 \hspace{2mm}
\includegraphics[angle=-90, width=0.95\textwidth]{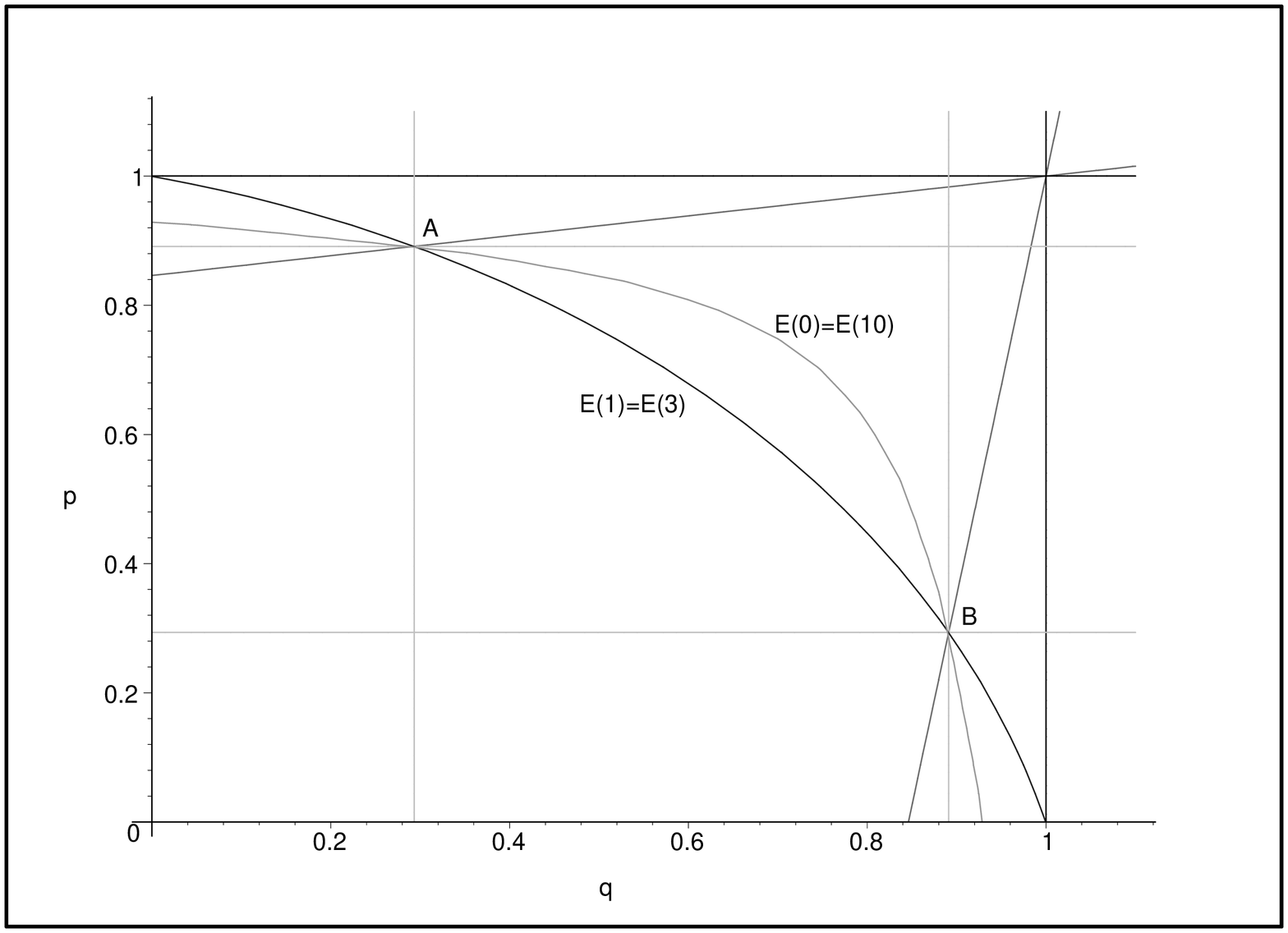}
 \end{minipage}
 \hspace{6mm}
 \begin{minipage}{0.45\textwidth}
 \includegraphics[angle=-90, width=0.95\textwidth]{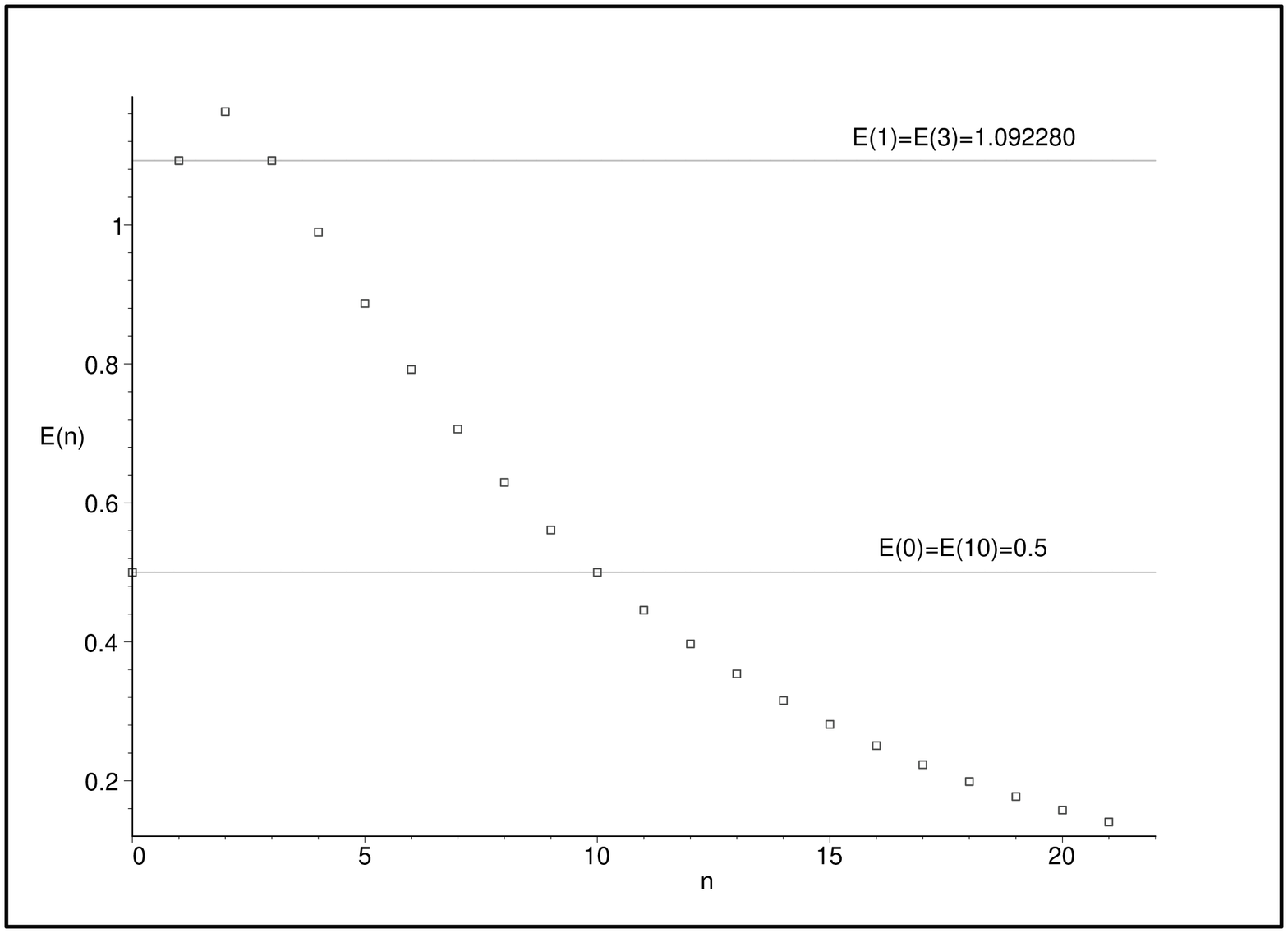}
 \hspace{-10mm}
\end{minipage}
\caption{{\em Left}: Intersection of the curves $E_{10}-E_0=0$ and $E_3-E_1=0$
at the points A($q_1$,$p_1$) and B($q_2$,$p_2$), see \eqref{eq21} for numerical values,
provides simultaneous degeneracies within the respective pairs of energy levels.
{\em Right}: The shape of energy spectrum $E_n=E_{q,p}(n)$ of
$q,\!p$-oscillator with $q_1$, $p_1$ or $q_2$, $p_2$ taken from
\eqref{eq21}. As seen, $E_0=E_{10}$ and $E_1=E_3$. }\label{fig5}
\end{figure}

The coordinates of the points where the curves $E_{10}-E_0=0$ and
$E_3-E_1=0$ mutually intersect are found from the equations
\begin{gather}\label{eq20}
E_{10}=E_0, \qquad E_3=E_1 .
\end{gather}
For the two intersection points (A) and (B) this yields:
\begin{gather}\label{eq21}
{\rm (A)} \ \  q_1=0.567239, \quad p_1=0.823554;  \qquad {\rm (B)} \ \
q_2=0.823554, \quad p_2=0.567239.
\end{gather}
Therefore, for either the pair (A) or the pair (B) in \eqref{eq21}, the
$q,\!p$-oscillator shows two double degeneracies from \eqref{eq20}. For
more clearness we illustrate this fact graphically in Fig.~\ref{fig5} (left).
Besides, for obtaining one-parameter $q$-oscillator, it is now
enough to take the relation $p=f(q)$ as linear function:
$   
p=\alpha q + \beta$,
with $\alpha$ and $\beta$ being such that the straight line passes
through the point $(1,1)$ and any one of the intersection points (A or
B) of the considered curves, see Fig.~\ref{fig5} (left). The parameters
$\alpha$ and $\beta$ satisfy one of the two systems of equations
     \begin{gather*}
\left \{
\begin{array}{l}
 \alpha q_1 + \beta  =  p_1  , \\
 \alpha + \beta = 1  ,
 \end{array}
\right.
\qquad
\left \{
 \begin{array}{l}
 \alpha q_2 + \beta = p_2  , \\
 \alpha + \beta = 1  ,
 \end{array}
\right.
\end{gather*}
so that the result is, respectively,
\begin{gather}\label{eq23}
{\rm (A)} \  \leftrightarrow  \ \alpha=0.407722, \
\beta=0.592278; \qquad {\rm (B)}  \ \leftrightarrow \
\alpha=2.452649, \ \beta=-1.452649.
\end{gather}
Substitution of the values from \eqref{eq21},  either the f\/irst pair or the
second, into the formula \eqref{eq7} for the energy $E_n=E_{q,p}(n)$ yields
the actual shape of energy function, as shown in Fig.~\ref{fig5} (right).
Note the monotonicity of energy function of one-parameter deformed
oscillator got from the linear relation.
  If we exploit the relation $p=\alpha q+ \beta$ with $\alpha$ and $\beta$
of the case (A) in \eqref{eq23}, i.e.\ the line passing through $(1,1)$ and the
point (A), we get the $q$-oscillator by excluding the parameter $p$,
or get the $p$-oscillator by excluding $q$ where $q=\alpha ^{-1}
(p-\beta)$. Though being on equal footing, the two versions of
one-parameter oscillator dif\/fer in the range of possible values of
$q$ versus those of $p$: indeed, $q \in (0,1]$ for the
$q$-oscillator whereas $p \in [p_{\min},1]$, $p_{\min}=\beta$, for the
$p$-oscillator. The situation which arises if instead of (A) one
uses the case (B) in \eqref{eq23}, is completely analogous to the previous
case of (A): this follows from the symmetry $q \leftrightarrow p$.

\section{Conclusions}\label{section5}

By explicit treatment, we have conf\/irmed that peculiarities of the
energy spectrum of $q$-deformed oscillator crucially depend on both
the particular version of $q$-oscillator and, within each version,
on the specif\/ied value of deformation parameter. As it was shown in
\cite{GR1}, the special TD (Tamm--Dancof\/f) version of
$q$-deformed oscillator admits the possibility of appearance, for
proper $q$, of the ``accidental'' double degeneracy within any one
chosen pair of energy levels. In the present paper, we have
demonstrated how to obtain one-parameter $p$-deformed oscillators
with the property of simultaneous degeneracy within each of the {\it
two} specif\/ied pairs of energy levels.
  Having chosen two pairs of energy levels, in particular, those given
in \eqref{eq10}, we applied dif\/ferent cases of quadratic relation $p=f(q)$:
parabolic, hyperbolic and elliptic ones, to the two-parametric
$q,\!p$-oscillator.
  On the other hand, with the specif\/ied parabolic dependence
we treated and obtained the cases of $p$-oscillator such that the
pairs of energy levels \eqref{eq18} or \eqref{eq19} exhibit the two-fold double
degeneracy.

We have also found that two double degeneracies can be achieved in
the general two-parameter case, without imposing any particular
relation $p=f(q)$, but only if the two pairs of energy levels belong
to some more special subset.
  The typical example, $E_0=E_{10}$ and $E_1=E_3$ in \eqref{eq20}, has been
considered and shown to provide the intersection points A$(q_1,p_1)$
and B$(q_2,p_2)$ in the $(q,p)$ plane which lead, in the
two-parameter case, to the desired two double (pairwise)
degeneracies.
  These same two degeneracies $E_0=E_{10}$ and $E_1=E_3$ do hold
for some one-parameter reduction in which the $q$-oscillator is
drawn by imposing linear relation $p=\alpha q + \beta$.

The obtained energy spectra are mainly given by non-monotonic
function: as seen from Fig.~\ref{fig1} (right), Fig.~\ref{fig3} (right), Fig.~\ref{fig4}
(right), in each of these cases the energy function $E_p=E_p(n)$
consists of two distinct parts, and this is rooted in the ``$ \pm $''
ambiguity of the inversion $q=f^{-1}(p)$, see \eqref{eq14} for example.
  In contrast, the situation described in Section~\ref{section4.2} results in the
monotonic energy function $E_p(n)$ since there is no ambiguity in
inverting the linear dependence: $q=f^{-1}(p)$.

It is worth to point out that in our paper we have in fact also
encountered the case of three-fold energy level degeneracy, see
Fig.~\ref{fig3} (right), where the energies $E_0$, $E_2$ and $E_5$ do
coincide. Clearly, the proposed approach to the deformed
$q,\!p$-oscillators and their various one-parameter reductions is
suitable for exploration of the degeneracies of energy levels
occurring in more complicated patterns such as, e.g., three double
(pairwise) degeneracies and others.

Let us make a remark concerning the ``no-go'' theorem mentioned in the
Introduction. The fact that $q,\!p$-oscillators, their special
$p = q$ case known as the TD oscillator, and the new deformed
oscillator models treated in this paper, really circumvent the
theorem and exhibit various patterns of ``accidental'' degeneracies,
is well explained by more involved structure of those
quantum-mechanical systems where relevant features should appear:
position dependent mass, the potential being a function of both
coordinate and momentum, etc.\ (see e.g.,~\cite{Mizr,QT,Wess}).

It would be of interest to analyze possible peculiarities of
applying, along the lines similar to those in \cite{G,AGI,AGI`}, the
respective versions of $q$-Bose gas model based on the above
constructed novel, nonstandard $q$-oscillators, in order to examine
their ef\/f\/iciency for description of the non-Bose type features shown
by the data on pion correlations collected in experiments on
relativistic heavy ion collisions. Certainly, other applications of
the exotic deformed oscillators treated in this paper are worth of
detailed study.

\subsection*{Acknowledgements}

This work was done under partial support of the Grant Number 14.01/016 of the
State Foundation of Fundamental Research of Ukraine.

\pdfbookmark[1]{References}{ref}
      \LastPageEnding

\end{document}